\documentclass[aps,pra,reprint,
nofootinbib,
nobibnotes,
amsmath,amssymb,
aps,amsthm,
]{revtex4-2}
\usepackage{hyperref}

\begin{document}
	\title{Objective Collapse Equation Maintains Conservation Laws With No New Constants}
	\author{Edward J. Gillis}
	\email[]{gillise@provide.net}

	\begin{abstract}

		\noindent 
		
		Modified versions of the Schr\"{o}dinger equation have been proposed in order to incorporate the description of measurement processes into the mathematical structure of quantum theory. Typically, these proposals introduce new physical constants, and imply small violations of momentum and energy conservation.

		These problematic features can be eliminated by assuming that wave function collapse is induced by the individual interactions that establish correlations between  systems. The generation of a sufficient number of small, random shifts of amplitude between interacting and noninteracting branches of the wave function can bring about collapse on a scale consistent with our macroscopic experience. Two-particle interaction potential energies can be used as the basis for a collapse term added to the Schr\"{o}dinger equation. The range of the interactions sets the distance scale of the collapse effects; the ratio of potential energies to the total relativistic energies of the particles determines the magnitude of the amplitude shifts, and the rate at which the interactions proceed fixes the timing parameters. 
		
		Consistency with conservation laws in individual experiments is maintained because the collapse operator automatically takes into account the small, residual entanglement between the measured system and systems with which it has previously exchanged conserved quantities during interactions. Conservation is exact for momentum and orbital angular momentum, and it holds for energy within the accuracy allowed by the limited forms of energy describable in nonrelativistic theory.

	\end{abstract}

	
	\maketitle

 \section{Introduction}

The main assumption made here is that wave function collapse is a real, nonlocal process that is induced by the kinds of \textit{individual} correlating interactions that constitute measurements. Collapse is achieved by shifting small amounts of amplitude between various components of the wave function until all amplitude is transferred either into or out of the interacting component(s). This idea can be implemented by adding stochastic, nonlinear terms to the nonrelativistic Schr\"{o}dinger equation. Nonlinearity is required in order to reproduce Born's rule as recently shown by Mertens, et al.\cite{Mertens_Born}. The need for a nonlinear equation to be stochastic was demonstrated by Gisin who showed that any deterministic nonlinear modification would lead to superluminal signaling\cite{Gisin_c}.

A number of equations of this kind have been proposed in order to explain measurement outcomes in terms of fundamental physical processes\cite{Pearle_1976,Pearle_1979,Gisin_1984,GRW,Diosi_1,Diosi_2,Diosi_3,GPR,Adler_Brun, Ghirardi_Bassi,Pearle_1,Brody_finite,Bassi}. In these accounts wave function collapse and the Born probability rule are
consequences of the basic mathematical structure of the theory, rather than ad hoc postulates, seemingly at odds with that structure.

The modified equations that have been proposed previously include new physical constants in order to determine the range, rate, and strength of the effects that generate collapse. These constants are fine-tuned to reproduce the predictions of conventional quantum theory at the microscopic level, while leading to collapse for macroscopic systems, and to minimize (but not eliminate) violations of conservation laws. 

The collapse equation described in Section 2 shares several of the basic structural features of prior works, but there are some key  differences. (a) The current proposal postulates a single, global stochastic process which is a function only of time, $\xi(t)$, in contrast to the assumption of a stochastic \textit{field} which is a function of position and time, $\xi(x,t)$. (b) Here the timing of collapse effects depends on the rate at which conserved quantities are exchanged during interactions between systems, rather than on a single arbitrary new constant. (c) The range of collapse effects depends on the range of the interaction potential energies, rather than on a second arbitrary new constant. (d) The only new ontological feature is the single global stochastic process; there is no need to posit either a mass or flash ontology. (e) The magnitude of changes of amplitude of the wave function is determined by the ratio of interaction energy to total relativistic energy - not by the combination of new constants and additional ontological features. (f) Finally, many proposals model the process of measurement with a simple, idealized Hamiltonian that couples the measured system to the apparatus, and assume that there is, at least very briefly, an actual superposition of distinct macroscopic states of the instrument. In contrast, here it is the individual, correlating interactions \textit{within} the apparatus that are primarily responsible for inducing collapse.

The proposed equation is described in the next section. Section 3 contains a proof that the equation is effective in inducing collapse in accord with the Born probability rule, and gives a rough estimate of the number of elementary interactions required to complete the collapse process. Section 4 shows how conservation laws are maintained in individual situations. Section 5 provides an estimate of the magnitude of predicted nonlinear effects and deviations from strict conservation of nonrelativistic forms of energy. Section 6 examines the most critical issue in constructing a relativistic version of the proposal - that of maintaining Lorentz invariance in an account of nonlocal collapse. Section 7 summarizes the argument.

\section{The Collapse Equation}

We begin with the nonrelativistic Schr\"{o}dinger equation in the form:
\begin{equation}\label{1}    
d\psi \,   = \,  (-i/\hbar)\mathbf{\hat{H}} \, \psi \, dt ,  
\end{equation}
and add stochastic nonlinear collapse terms:
\begin{equation}\label{2}   
\begin{array}{ll}  
d\psi \,   = \,  (-i/\hbar)\mathbf{\hat{H}} \, \psi \, dt \,  +\, 
\sum_{j < k} \hat{\mathcal{V}}_{jk}   \, \, \psi\, d\xi(t)  
& \\   
\; \; \;\;\;\; - \, \frac{1}{2}  (\sum_{j < k} \hat{\mathcal{V}}_{jk}) \,  
(\sum_{m < n} \hat{\mathcal{V}}_{mn} ) \, \psi \, dt .  
\end{array}
\end{equation}

The modified equation is interpreted according to the It$\hat{o}$ stochastic calculus rules.\cite{Ito_51} $d\xi(t)$ is a white noise Gaussian process whose integral, $\xi(t)$, is a Wiener process with zero mean. The differentials, $d\xi(t)$, obey the It$\hat{o}$ rules,    
$ d\xi^{*} d\xi = dt, \;\;\; dt d\xi = 0, $ and these determine the units, 
$ d\xi \,= \, (dt)^{\frac{1}{2}}.$ The middle term on the right hand side of equation \ref{2} is based on two-particle interaction potential energies, $\mathbf{\hat{ V}_{jk}}$, between system $j$ and system $k$. It is primarily responsible for shifting amplitude between components. The third term on the right hand side, 
$- \, \frac{1}{2}  (\sum_{j < k} \hat{\mathcal{V}}_{jk}) \,  
(\sum_{m < n} \hat{\mathcal{V}}_{mn} ) \, \psi \, dt $, preserves the norm of the wave function.

The interaction potential, 
$\mathbf{\hat{ V}_{jk}} \, = \, \mathbf{\hat{ V}_{jk}} (r)$, depends only on the separation, $r$, between systems $j$ and $k$, and its magnitude decreases as $r$ increases. The full stochastic operator, $\hat{\mathcal{V}}_{jk}$, is constructed by subtracting the expectation value, $\langle \, \psi | \mathbf{\hat{ V}_{jk}}| \psi \, \rangle$, from $\mathbf{\hat{ V}_{jk}}$, dividing the result by the sum of the relativistic energies of systems $j$ and $k$, and multiplying by $\sqrt{\gamma_{jk}}$ where $\gamma_{jk}$ is a parameter based on the rate at which the interaction proceeds:  
\begin{equation}\label{3}    
\hat{\mathcal{V}}_{jk} \; \equiv \;  \sqrt{\gamma_{jk}}\Big{[}(\mathbf{\hat{ V}_{jk}} - \langle \, \psi | \mathbf{\hat{ V}_{jk}}| \psi \, \rangle) \Big{]} \; \Big{ / } \; \Big{[}(m_j+m_k)c^2\Big{]}.
\end{equation}
The mass terms in the denominator reflect the extent to which the interaction can alter the state of the systems involved, and, thus, the degree of correlation that is established between them. These terms should be understood as representing the \textit{effective} mass; this can be done by aggregating  particles in bound states such as atoms, molecules, and other complex structures into single systems with a total mass and a net charge (or electric multipole moment).

The strength of the stochastic effect of individual interactions is determined by the ratio of energies in the collapse operators. The speed of light, $c$, is chosen to set this ratio because it is the only nonarbitrary speed. Because this proposal is formulated in a nonrelativistic framework, it can be interpreted as the speed of light in the preferred reference frame. The nonrelativistic limit on interaction energies implies that the ratio is less than about $10^{-3}$.  The relevant potentials are electrostatic, generally with a $1/r$ distance dependence.

The rate parameter for each interacting pair of systems,  $\gamma_{jk}$, is defined as the ratio of the magnitude of the rate of change of potential energy to the maximum possible change in potential energy expected during the interaction. In order to insure compliance with the Born probability rule these measures are restricted to the interacting components of the two systems, $j$ and $k$. These components are picked out by multiplying the wave function $\psi$ by the ratio, 
$ [ \, \mathbf{\hat{ V}_{jk}} \, / \,\langle \,\psi | \mathbf{\hat{ V}_{jk}}| \psi \, \rangle \, ] $: 
\begin{equation}\label{4}
\psi_{jk} \; \equiv \; [ \, \mathbf{\hat{ V}_{jk}} \, / \, \langle \, \psi | \mathbf{\hat{ V}_{jk}}| \psi \, \rangle \, ] \, \psi.
\end{equation}
\begin{equation}\label{5} 
\gamma_{jk} \; \equiv \; 
\frac { || \, \int \,  d/dt(\psi_{jk}^* \, \psi_{jk}  \, \mathbf{\hat{ V}_{jk}} ) \,   || }  { || \, \int \,  (\psi_{jk}^* \, \psi_{jk}  \, \mathbf{\hat{ V}_{max}} )\, || }.
\end{equation}
The numerator in equation \ref{5} can be computed as the magnitude of the rate of change of kinetic energy in the center-of-mass frame of the two interacting components:
\begin{equation}\label{6} 
\begin{array}{ll}            
\Big{|} \Big{|} \; \; \int \, d/dt( \psi_{jk}^* \, \psi_{jk} \mathbf{\hat{ V}_{jk}}  )   \;   \Big{|} \Big{|} \;  
& \\   
&  \\
\; \; \; \;  =  \; \Big{|} \Big{|} \;  (i \hbar ) \,
\int  \;  \Big{[} \,  \psi_{jk}^* \, \psi_{jk}
\, \Big{(} \,
\frac{\mathbf{\nabla_j}^2   \mathbf{\hat{ V}_{jk}} } {2m_j} \, 
+ \, \frac{\mathbf{\nabla_k}^2   \mathbf{\hat{ V}_{jk}} } {2m_k} \, \Big{)} 
& \\   
\;\;\;\; \;\;\;\; \; \; +  \;  \psi_{jk}^* \, \,
\mathbf{\hat{\nabla}_j}\mathbf{\hat{ V}_{jk}} \,  \cdot \,    \Big{(} \, \frac{\mathbf{\nabla_j} \psi_{jk}}{m_j}  \, - \, 
\frac{\mathbf{\nabla_k} \psi_{jk}}{m_k} \,  \Big{)} \,  \Big{]} \; \Big{|} \Big{|}.
\end{array}
\end{equation}
These integrals  and those in equation \ref{7} below are over the entire configuration space. The denominator in equation \ref{5} depends on the specific configuration of the wave function and the sign of the potential. For positive potentials it is the sum of the potential energy and the radial component of the kinetic energy; for negative potentials it is the norm of the effective potential of the lowest available energy state:
\begin{equation}\label{7}  
\begin{array}{ll}  
|| \, \langle \, \psi_{jk}^* \, | \, \mathbf{\hat{ V}_{max}}  \, | \, \psi_{jk} \, \rangle \,  ||_{pos}   \; 
& \\   
\;\;\;\;\;\;\;\;\;\;\; \equiv \; 
\int \, \psi_{jk}^* \, \Big{[} \, \mathbf{\hat{ V}_{jk}} \psi_{jk} \, - \, (\frac{\hbar^2}{\mu_{jk}}) \partial^2 \, \psi_{jk} / \partial r^2  \,  \Big{]},
& \\  
|| \, \langle \, \psi_{jk}^* \, | \, \mathbf{\hat{ V}_{max}}  \, | \, \psi_{jk} \, \rangle \,  ||_{neg}   \; 
& \\   
\;\;\;\;\;\;\;\;\;\;\;  \equiv \; 
|| \, \mathbf{\hat{V}_{jk}}' \, + \,  \mathbf{\hat{L}_{jk}}^2 /r^2 \,||  .
\end{array}
\end{equation} 
In the expression for positive potentials,  $ \mu_{jk} $ is the reduced mass of systems $j$ and $k$, $ (m_j * m_k) / (m_j + m_k)$; for negative potentials $ \mathbf{\hat{L}_{jk}}$ is the orbital angular momentum calculated in the $j-k$ center-of-mass frame, and the prime indicates the lowest available state.

The rate parameter is designed so that it is only effective during the brief period during which the correlations between the two systems are being established through the exchange of conserved quantities. It can change during this period due to interactions with systems other than $j$ and $k$, but for those interactions that are most important to inducing collapse the term, $\gamma_{jk} dt$, integrates to a value of order $1$ over the course of the interaction, and the same is true for the term, $\sqrt{\gamma_{jk}} d\xi(t).$ This implies that the overall effect of a single interaction is to transfer amplitude into or out of the interacting component of the wave function by an amount roughly equal to the maximum value of the ratio, $[(\mathbf{\hat{ V}_{jk}} - \langle \, \psi | \mathbf{\hat{ V}_{jk}}| \psi \, \rangle)] \; / \; [(m_j+m_k)c^2]$, that occurs during the interaction. From the interaction term in the Schr\"{o}dinger equation, 
\begin{equation}\label{8}    
d\psi \,   = \,  \frac{-i \mathbf{\hat{ V}_{jk}} } {\hbar} 
\, \psi \, dt ,  
\end{equation}
we can define a characteristic interaction time, 
\begin{equation}\label{9}    
dt_{int} \,  \equiv \, \frac{\hbar} {  \Delta \mathbf{\hat{ V}_{jk}} },   
\end{equation}
where $ \Delta \mathbf{\hat{ V}_{jk}} $ is the total change in $ \mathbf{\hat{ V}_{jk}} $ over the course of the interaction. This is about $10^{-18}$ to $10^{-17}$ seconds for the strongest nonrelativistic interactions.

\section{Measurement, Collapse, and the Born Rule}

The collapse process implied by equation \ref{2} can be described by tracing the evolution of the wave function in configuration space under the assumption that the total system consists of a very large number of subsystems that have been interacting for an extended period. 

A measurement of a particular quantity typically begins by physically separating components of the target system based on the value of the quantity in question. This physical separation is associated with the branching of the wave function of the entire system into distinct, nonoverlapping regions of configuration space.\footnote{Although the amplitude is not strictly zero between the branches, the collapse operator cleanly separates the wave function into orthogonal components with positive and negative values of $\hat{\mathcal{V}}_{jk}$.} Some or all of the components of the target system then undergo further interactions that effectively prevent the branches from recombining. At this stage the Hamiltonian acts in an essentially independent manner on each branch. Subsequent interactions bring about  a final  state of the systems involved that is macroscopically distinct from the initial state.

The primary effect of the stochastic operator, $\hat{\mathcal{V}}_{jk}$, associated with each \textit{individual} correlating interaction is to shift amplitude between the interacting branch and the remainder of the wave function. Hence, its action is essentially independent of that of the Hamiltonian. Since $\frac{\hat{\mathbf{V}}_{jk}} { (m_j+m_k)c^2 } \;   \leq \; 10^{-3}$ the variations induced by $\hat{\mathcal{V}}_{jk}$ are small perturbations on the ordinary Schr\"{o}dinger evolution within the interacting branch and essentially zero on other branches.

To trace the collapse process in detail we begin by calculating the change in $\psi^* \psi$ at each point in configuration space:   
\begin{equation}\label{10} 
\begin{array}{ll} 
d(\psi^*\psi) \; = \; (\psi^* + d\psi^*)(\psi + d\psi) \, - \, (\psi^*\psi) 
& \\  
\;\;\;\;\;\;\;\;\;  = \;   d\psi^* \, \psi \, 
+ \psi^* \, d\psi \,  +  d\psi^* \, d\psi. 
\end{array}
\end{equation}
(The It$\hat{o}$ calculus rules require retaining terms that are quadratic in the differentials $ d\psi^* \, d\psi  $ since 
$ d\xi^* \, d\xi \; = \; dt. $)
By using equation \ref{2} and its adjoint, and noting that the terms that are quadratic in $\sum_{j < k} \hat{\mathcal{V}}_{jk}$ cancel, the different effects of the Hamiltonian and the stochastic operator can be explicitly displayed: 
\begin{equation}\label{11}  
\begin{array}{ll}
d(\psi^*\psi) \; = \;  
\frac{i}{\hbar} \Big{[} \, (\mathbf{\hat{H}} \, \psi^*) \psi 
\, - \, \psi^* (\mathbf{\hat{H}} \psi) \, \Big{]} \, dt    
& \\    
\; \;\; \;   \; \;\; \; \; \;  \; \;\; \;  \;\; \;\; + \; \; 
(\psi^* \, \sum_{j < k} \hat{\mathcal{V}}_{jk} \, \psi)  \, (d\xi^* \, + \,  d\xi). 
\end{array}
\end{equation}
As stated, these two terms act in an essentially independent manner.

Individual interaction terms, $\hat{\mathcal{V}}_{jk}$, expand to:
\begin{equation}\label{12}  
\frac{ \Big{ ( } \, \mathbf{\hat{ V}_{jk}} - \langle \, \mathbf{\hat{ V}_{jk}} \, \rangle \, \Big{ ) } \,\sqrt{\gamma_{jk}}  } 
{  (m_j+m_k)c^2  }.  
\end{equation}
Designate the interacting component as $I$, the noninteracting, orthogonal component(s) as $O$, and the average value of $\mathbf{\hat{ V}_{jk}}$ over $I$ as:  
\begin{equation}\label{13}  
\overline{\mathbf{ V}^I_{jk}} \; \equiv \;  
\frac{  \langle \, I \, |V_{jk}| \, I \, \rangle  } 
{  \langle \, I \, | \,  I \, \rangle  }  
\end{equation}
Label the value of the integral of $\psi^* \psi$ over the interacting branch as $\mu^* \mu$, and that of the orthogonal branch,  
$\nu^* \nu \, = \, 1 \, - \, \mu^* \mu$. The average of $\mathbf{\hat{ V}_{jk}}$ over the entire wave function is then: $ \langle \, \mathbf{\hat{ V}_{jk}} \, \rangle \; = \, \mu^*\mu \, 
\overline{\mathbf{ V}^I_{jk}} $. The average values of $(\mathbf{\hat{ V}_{jk}} - \langle \, \mathbf{\hat{ V}_{jk}} \, \rangle) $ across the interacting  component and noninteracting components are:  
\begin{equation}\label{14}  
(int): \;
\overline{\mathbf{ V}^I_{jk}} \, (1 \, -  \mu^*\mu \, ) 
\, = \, \nu^* \nu \, \overline{\mathbf{ V}^I_{jk}}; 
\;\;\; (nonint): \;
-  \mu^* \mu \, \overline{\mathbf{ V}^I_{jk}}.
\end{equation} 
The total value of $ \hat{\mathcal{V}}_{jk} $ over each of these components is then: 
\begin{equation}\label{15}   
\begin{array}{ll} 
\; \; \; \;  \nu^* \nu \,   
\langle \, I \, | \,  I \, \rangle \;  
\,  \Big{ [ } \,\psi^* \; 
\frac{\overline{\mathbf{ V}^I_{jk}} \,\sqrt{\gamma_{jk}} \, }
{ (m_j+m_k)c^2 } \, \; \psi \, \Big{ ] }  
& \\   & \\  
\;\;\;\;\;\; = \; 
+ \, (\mu^* \mu \, \nu^* \nu) \,   \,  \Big{ [ } \,\psi^* \; 
\frac{\overline{\mathbf{ V}^I_{jk}} \,\sqrt{\gamma_{jk}} \, }
{ (m_j+m_k)c^2 } \, \; \psi \, \Big{ ] }  ; 
\; 
& \\    & \\   
- \, \mu^* \mu \, \langle \, O \, | \,  O \, \rangle \;  
\Big{ [ } \,\psi^* \; 
\frac{\overline{\mathbf{ V}^I_{jk}} \,\sqrt{\gamma_{jk}} \, }
{ (m_j+m_k)c^2 } \, \; \psi \, \Big{ ] }  
& \\   & \\  
\;\;\;\;\;\; = \; 
- \,  (\mu^* \mu \, \nu^* \nu) \,   \,  \Big{ [ } \,\psi^* \; 
\frac{\overline{\mathbf{ V}^I_{jk}} \,\sqrt{\gamma_{jk}} \, }
{ (m_j+m_k)c^2 } \, \; \psi \, \Big{ ] } . \, 
\end{array}
\end{equation} 
So the effects of $ \hat{\mathcal{V}}_{jk} $ on the interacting and noninteracting branches are equal and opposite, and the process generated by the stochastic operator can be viewed as  an unbiased random walk of the integrated value, $\mu^* \mu$, between the end points, $0$ and $1$. 

With a large enough number of interactions all amplitude will be transferred either into or out of each interacting branch. 
The probability that simultaneous interactions in distinct branches could frustrate the collapse is vanishingly small since these interactions would have to be almost perfectly synchronized in both time and interaction strength over a walk of more than $10^6$ steps. The possibility of collapse to a single point is eliminated by the very short duration and relative weakness of the collapse operator relative to the Hamiltonian.

To establish conformity with the Born rule it is necessary to show that the probability that the squared amplitude, 
$\mu^* \mu$, of an interacting branch is eventually increased to $1$ is $\mu^* \mu$. The total change in the squared amplitude occurring at any time can be represented by using equation \ref{15} and summing over all of the interactions taking place in the interacting branch:  
\begin{equation}\label{16}
\pm \; (\mu^* \mu \, \nu^* \nu) \,  \mathbf{ \sum_{jk}} \,  \Big{ [ } \,\psi^* \; 
\frac{\overline{\mathbf{ V}^I_{jk}} \,\sqrt{\gamma_{jk}} \, }
{ (m_j+m_k)c^2 } \, \; \psi \, \Big{ ] }   \, (d\xi^*(t) \, + \,  d\xi(t)). 
\end{equation} 
Each individual term integrates to a value less than $10^{-3}$ over the course of the interaction. The stochastic process, $\xi(t)$ is continuous, and it has a variance of $t_2 - t_1$ over a time period from $t_1$ to $t_2$. By taking time increments, $dt$, that are sufficiently small one can insure that the sum in equation \ref{16} is less than $1$. Since 
$\mu^* \mu \; \nu^* \,  \nu$ is updated continuously the total value of the expression in equation \ref{16} is less than or equal to the difference between $\mu^* \mu $ and either end point ($0$ or $1$). Designate this value as $\delta$. Label the probability of $\mu^* \mu $ eventually reaching $1$ as $Pr(\mu^* \mu)$; so $Pr(0) \, = \, 0$, $Pr(1) \, = \, 1$. Since the step size, $\delta$, is less than $ \mu^* \mu \; \nu^* \nu $ it is less than or equal to the distance to either end point. Since steps in either direction are equally likely 
\begin{equation}\label{17}
Pr(\mu^* \mu) \, = \, \frac{1}{2}[ Pr(\mu^* \mu-\delta)] \, + \, \frac{1}{2}[ Pr(\mu^* \mu+\delta)] 
\end{equation}  
for all values of $\mu^* \mu $ and $\delta$. Therefore, $Pr(\mu^* \mu)$ is linear, and the Born rule follows. This insures consistency with relativity at the level of observation.

Since the magnitude of amplitude shifts is determined largely by the ratio of interaction energy to total relativistic energy, it is also possible to estimate the number of elementary correlating interactions involved in a typical collapse. The nonrelativistic limit on the ratio is roughly $10^{-3}$. In most measurement scenarios we would not expect the average ratio to be less than about $10^{-7}$. In a random walk of $N$ steps the standard deviation from the starting point is $\sqrt{N}$ times the average step size. If the step size were determined solely by the interaction ratios, then collapse should occur after approximately $10^6$ to $10^{14}$ interactions. However, as shown above the step size is also dependent on the product of the squared amplitudes, $\mu \mu^* * \nu \nu^*$, of the interacting and noninteracting components. As the process approaches either end point this product becomes quite small. If we take a value of $10^{-3}$ as an effective lower limit on this quantity then the length of the walk is extended to about $10^{12}$ to $10^{20}$ steps. This is still comfortably in the mesoscopic or very small macroscopic range. The duration of individual interactions is about $10^{-18}$ to $10^{-11}$ seconds. However, since the number of interactions taking place simultaneously would increase very sharply as the measurement interaction proceeds, collapse is likely to be complete in a small fraction of a second. 

These estimates apply to well organized measurements that take place in a laboratory. More disorganized collapse processes that occur in a natural setting could take longer.

\section{Conservation Laws} 

After an extended period of interaction nearly all systems are entangled to some extent\cite{Gemmer_Mahler,Durt_1}. Conserved quantities are \textit{shared among these systems}. The branching processes referred to above are brought about by conservative interactions that split \textit{both} interacting systems into distinct parts. The interaction alters the distribution of conserved quantities between the systems, but it does not alter the total amount of those quantities in either branch. This fact is readily appreciated when two elementary systems interact and become entangled, but it is often overlooked when one of the interacting systems is macroscopic. A photon passing through a beam-splitter provides a simple illustration of this point in the latter case. The exchange of momentum between the reflected branch and the beam-splitter changes the state of the reflecting component of the beam-splitter as well as that of the photon. There is no resulting difference in the total momentum of the two distinct branches.

Given the equality between branches strict compliance with conservation laws in individual experiments can be established by showing that equation \ref{2} implies that the (normalized) change in the relevant quantities within each branch is zero. This will be demonstrated exactly for momentum and angular momentum. A generally similar result will be shown for energy. The apparent violations of energy conservation that are sometimes attributed to the narrowing of the measured system's wave function are compensated for by correlated changes in entangled systems. However, when there is an exchange of conserved quantities between systems the standard nonrelativistic accounting does not accurately track energy changes associated with relativistic mass corrections, radiation, and small contributions from antiparticles. Since equation \ref{2} accommodates only the usual nonrelativistic types of energy it implies small deviations from strict conservation in these cases. These deviations closely parallel the discrepancies entailed by ordinary Schr\'{o}dinger evolution. In stationary states and noninteracting situations equation \ref{2} reduces to the Schr\"{o}dinger equation, and there are no deviations or discrepancies regarding changes in energy. This proposal is based on the premise that collapse occurs in precisely those situations in which the discrepancies occur. It is possible, therefore, that these deviations are an artifact of the nonrelativistic formulation of the theory, and that they could be eliminated by an extension of this proposal that fully accounts for relativistic effects.

Following the derivation of equation \ref{11} above and abbreviating $\sum_{j < k} \hat{\mathcal{V}}_{jk} \, \equiv \, \hat{\mathcal{V}}$, the change induced by an operator, $\mathbf{\hat{Q}}$, at each point in configuration space is: 
\begin{equation}\label{18}   
\begin{array}{ll} 
d (\psi^* \, \mathbf{\hat{Q}} \,\psi) \; = \; 
\frac{i}{\hbar} \Big{[} \, (\mathbf{\hat{H}} \, \mathbf{\hat{Q}} \,\psi^*) \psi 
\, - \, \psi^* (\mathbf{\hat{Q}} \,\mathbf{\hat{H}} \psi) \, \Big{]} \, dt  
\; 
& \\    
\; \;    
-\psi^* (\frac{1}{2}  \hat{\mathcal{V}}
\hat{\mathcal{V}}) \, \,\mathbf{\hat{Q}} \, \psi \, dt \; - \;
\psi^* \, \hat{\mathbf{Q}} \, (\frac{1}{2}\hat{\mathcal{V}}
\hat{\mathcal{V}}) \, \psi \, dt \,   
+ \; \psi^* \, \hat{\mathcal{V}}  \hat{\mathbf{Q}}  \hat{\mathcal{V}} \,  \psi   \, d\xi^*d\xi
\;  
& \\  
\; \;\; \;   \; \;\; \; \; \;  \; \;\; \;   \; \;\; \;   \;\;       
\; \; + \; \;  \psi^* \,  \hat{\mathcal{V}}  \hat{\mathbf{Q}} \, \psi  \, d\xi^* \; + \;\psi^* \,  \hat{\mathbf{Q}}   
\hat{\mathcal{V}} \, \psi  \, d\xi.   
\end{array}
\end{equation} 
The three quantities that are of interest here all commute with the Hamiltonian. For momentum and angular momentum this is insured by the restriction to two-particle, distance-dependent potentials, and it is obviously true for energy. So these quantities are conserved across the wave function, and since the Hamiltonian operates independently on each branch it also conserves the quantities within each branch. Therefore, to determine whether equation \ref {2} respects the relevant conservation laws, we need only to assess the effect of the stochastic operator:   
\begin{equation}\label{19}   
\begin{array}{ll} 
d (\psi^* \, \mathbf{\hat{Q}} \,\psi)_{col} \;          
\; = \;   \;  
-\frac{1}{2}\psi^* [(\hat{\mathcal{V}})^2 \mathbf{\hat{Q}}  \, + \,
\hat{\mathbf{Q}}  (\hat{\mathcal{V}})^2 ]\psi \, dt  
& \\  
\;  \; \;\; \;   \; \;\; \; \; \;  \; \;\; \;   \; \;\; \;   \;\;       
\; \; \;   \;\; \;   \;\; 
+ \; \psi^*( \hat{\mathcal{V}}  \hat{\mathbf{Q}}  \hat{\mathcal{V}} \, )\psi   \, dt \; \;   
& \\  
\; \;\; \;   \; \;\; \; \; \;  \; \;\; \;   \; \;\; \;   \;\;       
\; \; + \; \;  \psi^* \,  \hat{\mathcal{V}}  \hat{\mathbf{Q}} \, \psi  \, d\xi^* \; + \;\psi^* \,  \hat{\mathbf{Q}}   
\hat{\mathcal{V}} \, \psi  \, d\xi.   
\end{array}
\end{equation} 
What will be shown here is that the change in these quantities induced by the collapse equation is directly proportional to the change that is induced in the squared amplitude at that point:  
\begin{equation}\label{20}
d(\psi^*\mathbf{\hat{Q}}\psi) \, / \,  (\psi^*\mathbf{\hat{Q}}\psi)  \; = \; d(\psi^* \, \psi) \, /  \, (\psi^* \, \psi).
\end{equation}
It is because the change in amplitude at each point in configuration space affects every subsystem, all of which have been shaped by previous interactions, that the collapse process is able to maintain strict conservation.

The change in $\psi^* \psi$ due to the stochastic operator was shown by equation \ref{11} to be 
\newline 
$ d(\psi^*\psi) \; = \;  (\psi^* \, \hat{\mathcal{V}} \, \psi)  \, (d\xi^* \, + \,  d\xi).  $  So we need to show that 
\begin{equation}\label{21}
d(\psi^*\mathbf{\hat{Q}}\psi) \; = \; 
(\psi^* \, \hat{\mathcal{V}} \,\mathbf{\hat{Q}}\psi)  \, (d\xi^* \, + \,  d\xi).
\end{equation}
From equation \ref{19} this can be done by demonstrating that
\begin{equation}\label{22} 
\hat{\mathbf{Q}}   \hat{\mathcal{V}} \, \psi 
\; = \; \hat{\mathcal{V}} \, \hat{\mathbf{Q}} \psi 
\end{equation} 
at each point in configuration space.

The momentum and angular momentum operators are:
\begin{equation}\label{23}  
\begin{array}{ll}
\mathbf{\hat{P}} \, \equiv \,  -i \hbar \sum_j \,\mathbf{\hat{\nabla}_j}; 
& \\ 
& \\   
\mathbf{\hat{L}} = -i \hbar \sum_j \, \mathbf{w_j}   \mathbf{\times}      \mathbf{\hat{\nabla}_j}.
\end{array}
\end{equation}
The only spatial dependence in the collapse operator, $\hat{\mathcal{V}} $, is in the terms representing the conservative interaction potentials, 
$\mathbf{\hat{ V}_{jk}}$. Since 
$ \mathbf{\hat{\nabla}_j}\mathbf{\hat{ V}_{jk}} \, = \, -\mathbf{\hat{\nabla}_k}\mathbf{\hat{ V}_{jk}},  \, $ we get 
\begin{equation}\label{24}
\begin{array}{ll}
\mathbf{\hat{P}} (  \mathbf{\hat{ V}_{jk}} \, \psi ) \, = \,  (-i \hbar ) \, [ \, (\mathbf{\hat{\nabla}_j} \,   \mathbf{\hat{ V}_{jk}} ) \, \psi  \, + \,     \mathbf{\hat{ V}_{jk}}  \,   ( \mathbf{\hat{\nabla}_j} \, \psi )   
& \\ 
\;\;\;\;  \;\;\;\;\;\;\;\;\;\;\;\;\;\;\;\; \;\;\;\;\; \; \; + \; 
(\mathbf{\hat{\nabla}_k} \,   \mathbf{\hat{ V}_{jk}} ) \, \psi  \, + \,     \mathbf{\hat{ V}_{jk}}  \,   ( \mathbf{\hat{\nabla}_k} \, \psi ) \,   ] 
& \\ 
\;\;\;\;\;  = \; 
(-i \hbar ) \, [ \, \mathbf{\hat{ V}_{jk}}  \,   ( \mathbf{\hat{\nabla}_j} \, \psi )   
\, + \,  \mathbf{\hat{ V}_{jk}}  \,   ( \mathbf{\hat{\nabla}_k} \, \psi ) \,   ]   \, = \,  
\mathbf{\hat{ V}_{jk}} \,  (\mathbf{\hat{P}}  \, \psi ) 
\end{array} 
\end{equation}
Therefore, 
\begin{equation}\label{25}
d(\psi^*\mathbf{\hat{P}}\psi) \; = \; 
(\psi^* \, \hat{\mathcal{V}} \,\mathbf{\hat{P}}\psi)  \, (d\xi^* \, + \,  d\xi),
\end{equation}
and momentum is strictly conserved by the collapse equation.

Before proceeding there are two potential misunderstandings concerning momentum conservation that should be addressed. First, although the strict conservation of momentum allows the center of mass to evolve in a \textit{unitary} manner, this does not interfere with the nonunitary collapse process, because the description of this evolution requires that the center of mass is \textit{decoupled from the rest of the system}. Second,a very recent work by Donadi, et al.\cite{momentum_spread} shows that collapse implies an increase in the \textit{variance} of the momentum. This argument does not affect the conclusion reached here since momentum conservation is concerned with the \textit{mean} value.

An explicit expansion of the x-component of angular momentum for systems, $j$, and $k$, gives: 
\begin{equation}\label{26} 
\begin{array}{ll} 
[ \mathbf{\hat{L}_x}(\mathbf{w_j}) \, + \, \mathbf{\hat{L}_x}(\mathbf{w_k}) ] 
(\mathbf{\hat{V}_{jk}} \, \psi )
& \\     \; \;\; \;  \;\; \;   \;\; \;  \;
\;  = \; 
-i\hbar[  (y_j \, + \, y_k) (\partial{\mathbf{\hat{V}_{jk}}}/\partial{z_j} 
\, + \, \partial{\mathbf{\hat{V}_{jk}}}/\partial{z_k} )
& \\     \; \;\; \;  \;\; \;   \;\; \;  \; \;\; \;  \; \; \;  \; \; - \;   
(z_j \, + \, z_k) (   \partial{\mathbf{\hat{V}_{jk}}}/\partial{y_j}  
\, + \, \partial{\mathbf{\hat{V}_{jk}}}/\partial{y_k} )] \; \psi
& \\     \; \;\; \;  \;\; \;   \;\; \;  \; \;\; \;  \; \; \;  \; \; + \;    \mathbf{\hat{V}_{jk}} \, (\mathbf{\hat{L}_x} \, \psi).
\end{array} 
\end{equation}  
Since:   
\begin{equation}\label{27} 
\begin{array}{ll} 
(\partial{\mathbf{\hat{V}_{jk}}}/\partial{z_j}   \, + \, \partial{\mathbf{\hat{V}_{jk}}}/\partial{z_k} ) \; = \; 0; 
& \\    
(\partial{\mathbf{\hat{V}_{jk}}}/\partial{y_j}   \, + \, \partial{\mathbf{\hat{V}_{jk}}}/\partial{y_k} ) \; = \; 0,
\end{array}
\end{equation}  
we get:
\begin{equation}\label{28}
\mathbf{\hat{L}_x} (  \mathbf{\hat{ V}_{jk}} \, \psi )  \; = \;  
\mathbf{\hat{ V}_{jk}} \,  (\mathbf{\hat{L}_x}  \, \psi )  
\end{equation}
with similar relationships for the $y$ and $z$ components. So, the stochastic operator also strictly conserves orbital angular momentum at each point.

To evaluate the effect of the stochastic operator on energy we need to examine the relationship between   
$ \mathbf{\hat{H}} \, \hat{\mathcal{V}}_{jk} $ and $ \hat{\mathcal{V}}_{jk} \,\mathbf{\hat{H}}.$ Since it is clear that, at each point in configuration space, 
$\mathbf{\hat{V}} \hat{\mathcal{V}}  
\; = \; \hat{\mathcal{V}}  \mathbf{\hat{V}}, $ the only deviations from perfect conservation that can arise are those involving the kinetic energy terms. The action of the kinetic energy operator on the stochastic operator is:
\begin{equation}\label{29} 
\begin{array}{ll} 
(-\frac{\hbar^2}{2m} \mathbf{\nabla}^2) \, \hat{\mathcal{V}}_{jk}  
\; \; =   \; \;
\hat{\mathcal{V}}_{jk}
(-\frac{\hbar^2}{2m} (\mathbf{\nabla_j}^2  \, + \,  \mathbf{\nabla_k}^2) ) \; 
&  \\  \; \; \; \; \; \; \; \; \; \; \; \;  +  \; \;
(-\frac{\hbar^2}{2m}) \, \{ \, (\mathbf{\nabla_j}^2   \hat{\mathcal{V}}_{jk} \, + \, \mathbf{\nabla_k}^2   \hat{\mathcal{V}}_{jk} )  
&  \\  \; \; \; \; \; \; \; \; \; \; \; 
\; + \;   
2 ( \mathbf{\nabla_j}  \hat{\mathcal{V}}_{jk} \, \cdot \, 
\mathbf{\nabla_j}  
\, + \, \mathbf{\nabla_k} \hat{\mathcal{V}}_{jk} \, \cdot \,  \mathbf{\nabla_k})\}  .
\end{array}
\end{equation} 
Applying this to equation \ref{19} we get:
\begin{equation}\label{30} 
\begin{array}{ll}
d (\psi^* \, \mathbf{\hat{H}} \,  \hat{\mathcal{V}} \, \psi)_{col} \; =  \; \; 
\psi^* \, \hat{\mathcal{V}}  \mathbf{\hat{H}} \, \psi \, \, (d\xi^* \, + \, d\xi) \; 
& \\ 
&  \\  \; \; \; \; \; \;\; 
\;  -  \;
(\frac{\hbar^2}{2m}) \,  \psi^* \,  \{ \sum_{j<k} \, [  (\mathbf{\nabla_j}^2  \hat{\mathcal{V}}_{jk} \, + \, \mathbf{\nabla_k}^2  \hat{\mathcal{V}}_{jk}) \psi 
& \\   \; \; \; \; \; \;\; \;  \; \; \; \; \; \;\; \; \; \;
\; + \;   2 (\mathbf{\nabla_j}\hat{\mathcal{V}}_{jk}  \, \cdot \, 
\mathbf{\nabla_j}\psi  
\, + \, \mathbf{\nabla_k}\hat{\mathcal{V}}_{jk} \, \cdot \,  \mathbf{\nabla_k}\psi)]\} \, d\xi  
& \\
&  \\  \; \; \; \; \; \;\; 
\;  +  \;
(\frac{\hbar^2}{2m})  \psi^*   \psi  \,   
(\mathbf{\nabla} \hat{\mathcal{V}} \cdot  \mathbf{\nabla} \hat{\mathcal{V}} )   \, dt. 
\end{array}
\end{equation}
The first line gives the desired proportionality between $d(\psi^* \, \mathbf{\hat{H}} \,\psi)_{col}$ and $d(\psi^* \, \psi)_{col}$. This term highlights the usually small amount of entanglement between the measured system and those systems (such as the ``preparation apparatus") which is so often overlooked in discussions of the way in which measurements affect energy conservation.

The apparent deviations from perfect energy conservation are represented on the middle and last lines. They are relevant only when the interacting component of the wave function is selected since $\mathbf{\nabla} \hat{\mathcal{V}} \approx 0$ over the noninteracting component. They can be compared to the first two relativistic corrections to the kinetic energy formula:
\begin{equation}\label{31}
\begin{array}{ll}
\frac{mc^2} { \sqrt{1-(v^2/c^2)} } \; - \; mc^2 
& \\
\;\;\;\;\;\;\;\;\;\; 
\; = \; 
mc^2 \Big{ [ }\frac{1}{2}\frac{v^2}{c^2} \, + \, 
\frac{3}{2} \Big{(} \frac{1}{2}\frac{v^2}{c^2} \Big{)}^2 \, + \, 
\frac{5}{2} \Big{(} \frac{1}{2}\frac{v^2}{c^2} \Big{)}^3 \, + \, ... \Big{]} \; 
& \\
\;\;\;\;\;\;\;\;\;\;
\; = \; 
KE_{nr} \Big{[} 1 \, + \,  \frac{3}{2} \frac{ KE_{nr}}{mc^2} 
\, + \, \frac{5}{2}  \Big{(}  \frac{ KE_{nr}}{mc^2} \Big{)}^2 \, + \, ... \Big{]}, 
\end{array}
\end{equation} 
where $KE_{nr}$ is the nonrelativistic kinetic energy. Other adjustments to the nonrelativistic energy accounting  
include radiation equal to approximately
$ KE_{nr} \Big{[} \frac{v}{c} \frac{ KE_{nr}}{mc^2} \Big{]}$, and anti-particle content which contributes a fraction of about $ (\frac{ KE_{nr}}{mc^2})^2 $ to localized wave packets, $\psi^* \psi$\cite{Bjorken_Drell}.

The individual terms from the middle lines of equation \ref{30} can be rewritten as:\begin{equation}\label{32}
\begin{array}{ll} 
(\frac{ i \,\hbar}{2m}) \mathbf{\big{[}} \psi^* \, \psi    (\mathbf{\nabla_j}^2 \, \mathbf{\hat{ V}_{jk}}  ) 
\; + \;      
2 \psi^*(\mathbf{\nabla_j} \mathbf{\hat{ V}_{jk}}  \, \cdot \, 
\mathbf{\nabla_j}\psi) \mathbf{\big{]}} \, \, 
&  \\ 
\;\;\;\;\;  \;\;\;\;\;     \;
(\sqrt{\gamma_{jk} }\, d\xi) \, 
(\frac{ i \, \hbar}{ (m_j+m_k) c^2)} ).  
\end{array}
\end{equation}
The conversion of potential to kinetic energy during an interaction according to the Schr\"{o}dinger equation is: 
\begin{equation}\label{33}
(\frac{i \hbar}{2m}) \mathbf{\big{[}} \psi^* \, \psi    (\mathbf{\nabla_j}^2 \, \mathbf{\hat{ V}_{jk}}  ) 
\; + \;      
2 \psi^*(\mathbf{\nabla_j} \mathbf{\hat{ V}_{jk}}  \, \cdot  \,
\mathbf{\nabla_j}\psi)  \mathbf{\big{]}} \; dt. 
\end{equation}
Comparison of equations \ref{32} and \ref{33} illustrates that the deviations implied by equation \ref{2} arise in exactly the same situations in which the conventional theory fails to account for all of the changes in kinetic energy associated with the lowest order relativistic corrections.

To determine the ratio of the deviation implied by equation \ref{32} to the change in kinetic energy described by equation \ref{33} we can use the time defined by the expression, $ \frac{ \hbar}{ (m_j+m_k) c^2)}  $ (since the integral of $(\sqrt{\gamma_{jk} }\, d\xi) \, $ is of order $1$), and compare it to the characteristic interaction time from equation \ref{9},  
$ dt_{int} \, = \, \frac{ \hbar} { \Delta \mathbf{\hat{ V}_{jk}}    }  \, =  \,  \frac{ \hbar}{ \Delta  KE_{jk} }$. This yields a ratio of 
$   \frac{ \Delta KE_{nr}}{ (m_j+m_k) c^2)},$ closely matching the first order correction from equation \ref{31}.

A careful parsing of the individual terms in the expression on the last line of equation \ref{30} shows that it is reduced from the first order deviation by another factor of 
$\frac{KE_{nr}}{ (m_j+m_k) c^2)} = \frac{\hat{ V}_{jk}}{ (m_j+m_k) c^2)},$ similar to the second order relativistic correction for kinetic energy. The fact that it is strictly positive might raise some concerns, but these can be alleviated by noting that it is of the same magnitude as the contributions from anti-particle amplitudes generated during interactions, and several orders of magnitude smaller than radiative effects. It should also be noted that a relativistic characterization of interaction energies would include more than just scalar potentials.

The close qualitative and quantitative parallels between the deviations implied by equation \ref{2} and the discrepancies entailed by the nonrelativistic Schr\"{o}dinger equation suggest that the apparent violations of strict energy conservation are an artifact of the nonrelativistic formulation presented here. It is possible, therefore, that an extension of this collapse proposal that includes relativistic effects could be fully consistent with the major conservation laws.

\section{Experimental Prospects}

Because equation \ref{2} assumes that collapse effects are induced by interactions the nonlinear deviations that it implies can only be exhibited by changing the basis in which the interactions occur. Since this is the technique employed in quantum eraser experiments we can use an idealized, generic quantum eraser to estimate the magnitude of the predicted nonlinearities. The basic idea of these experiments is to divide the wave function of a ``target" particle into two separate branches, have one branch interact with a ``detector" particle, recombine the branches of the target particle, and then measure both target and detector in a basis that is complementary to that in which the interaction took place.

If we label the target and detector particles as $T$ and $D$, the interacting and noninteracting branches as $I$ and $O$, and the complementary symmetric and antisymmetric branches as $S$ and $A$, then a generic quantum eraser experiment can be represented schematically as: 
\begin{equation}\label{34}
\begin{array}{ll} 
(1/\sqrt 2)(|T_I\!\rangle + |T_O\!\rangle \;) \otimes|D\!\rangle  & \\   \Longrightarrow \;  
(1/\sqrt 2)(|T_I\!\rangle|D_I\!\rangle + |T_O\!\rangle |D_O\!\rangle  ) 
& \\  
= \; (1/\sqrt 2)(|T_S\!\rangle|D_S\!\rangle + |T_A\!\rangle |D_A\!\rangle  ).
\end{array}
\end{equation}
The symmetric and antisymmetric states are defined as:
\begin{equation}\label{35}
\begin{array}{ll} 
|T_S\!\rangle  =  (1/\sqrt 2)(|T_I\!\rangle + |T_O\!\rangle \;) \, ; 
\;\;\; \;\;\; 
|T_A\!\rangle  =  (1/\sqrt 2)(|T_I\!\rangle - |T_O\!\rangle \;) 
& \\   
|D_S\!\rangle  =  (1/\sqrt 2)(|D_I\!\rangle + |D_O\!\rangle \;) \, ; 
\;\;\; 
|D_A\!\rangle  =  (1/\sqrt 2)(|D_I\!\rangle - |D_O\!\rangle \;). 
\end{array}
\end{equation}
If the evolution is perfectly linear then measurements made in the symmetric-antisymmetric basis would yield a perfect correlation between the states of the target and detector particles (assuming that the amplitudes of the initial $I$ and $O$ branches were equal). Either both would be $S$ or both would be $A$. Equation \ref{2} predicts deviations from this perfect correlation dependent on the interaction energy. These can be estimated as follows. 

The principal stochastic term in equation \ref{2} is 
$ \sum_{j < k} \hat{\mathcal{V}}_{jk}   \, \, \psi\, d\xi(t). $ (The normalization term can be neglected because it is higher order in a small quantity.) We can focus on the deviation induced by a single complete interaction between systems, $j$ and $k$, $ \hat{\mathcal{V}}_{jk}   \, \, \psi\, d\xi(t)$. Individual operators were defined as: 
\newline 
$  \hat{\mathcal{V}}_{jk} \; \equiv \; \frac{ (\mathbf{\hat{ V}_{jk}} - \langle \, \psi | \mathbf{\hat{ V}_{jk}}| \psi \, \rangle)}{(m_j+m_k)c^2)}
\, \sqrt{\gamma_{jk}}. $
Over the course of the interaction the term, $ \sqrt{\gamma_{jk}}  d\xi(t)$, integrates to a value of order, $1$. If we abbreviate the average value of $  \frac{ (\mathbf{\hat{ V}_{jk}} - \langle \, \psi | \mathbf{\hat{ V}_{jk}}| \psi \, \rangle)}{(m_j+m_k)c^2)}  $ over the course of the interaction as $\epsilon$ then the principal stochastic effect of the initiating interaction in a quantum eraser experiment can be represented as: 
\begin{equation}\label{36}   
\begin{array}{ll}  
\hat{\mathcal{V}}_{jk}   \, \, \psi\, d\xi(t)  \; = \; 
\frac{1}{\sqrt 2} ( \pm \epsilon |T_I\!\rangle|D_I\!\rangle \mp \epsilon  |T_O\!\rangle |D_O\!\rangle  ) 
& \\  
= \; \pm \, \frac{\epsilon}{2 \sqrt 2}[ \; (|T_S\!\rangle + |T_A\!\rangle) \otimes 
(|D_S\!\rangle + |D_A\!\rangle) & \\  
\;\;\; \;\;\;\; \; \;  - \;\;\; \; (|T_S\!\rangle - |T_A\!\rangle) \otimes  (|D_S\!\rangle - |D_A\!\rangle) \; ] 
& \\  
= \; \pm \, \frac{\epsilon}{\sqrt 2} \,   
[ \, |T_S\!\rangle|D_A\!\rangle +  |T_A\!\rangle |D_S\!\rangle \, ].
\end{array}
\end{equation}
These cross $S-A$ terms, which are completely absent under strictly linear evolution, would appear with an amplitude of order, 
$ \epsilon \, \sim \, \frac{ \mathbf{\hat{ V}_{jk}}} {(m_j+m_k)c^2)}.  $ So the probability of observing them in this idealized arrangement would be the square of this ratio of interaction energy to total relativistic energy. For nonrelativistic interactions this is less than $10^{-6}$. Detection of such effects would be extremely challenging.

The strategy of looking for small violations of energy conservation, which has been used to test other collapse proposals, is not a viable way of testing equation \ref{2}. The argument of the previous section suggests that there probably are not any such violations. Even if the possible violations discussed in the previous section are not just artifacts of the nonrelativistic formulation their magnitude would be extremely small for any systems that could reasonably be studied. In this proposal wave function collapse is tied to correlating interactions. The number of such interactions and the interaction energy are functions of the thermal energy of a system.  A rough estimate of the magnitude of possible violations in terms of the thermal energy can be developed as follows.

The thermal energy of a system depends on the number of particles, the temperature, and Boltzmann's constant, $ \sim \; NkT$. The rate at which interactions involving energy exchange occur, $X$, depends on the thermal speed of the particles, $v$, and their separation, $d$: $X \sim v/d$. Since the interaction energy is $\sim \; kT$ the rate of \textit{possible} fractional increase of nonrelativistic energy described in Section 5 is approximately 
\begin{equation}\label{37}
\Big{(}\frac{kT}{mc^2} \Big{)}^2  \Big{(}\frac{v}{d} \Big{)}.
\end{equation}

Consider a volume of air at standard temperature and pressure. The average kinetic energy of the molecules is roughly $4*10^{-21}$ joules. The average mass is around $5*10^{-26}$ kilograms which means that $mc^2 \approx 4.5*10^{-9}$ joules. Thus the ratio of interaction energy to relativistic energy is of the order of $10^{-12}$. The square of this value is about $10^{-24}$. On average each molecule undergoes about $10^{10}$ collisions (interactions) per second, yielding a rate of fractional increase of about $10^{-14}$ per second. A cubic meter of air at standard temperature and pressure contains approximately $2.5*10^{25}$ molecules and a thermal energy of about $10^5$ joules. In one year ($ \sim \, 3.16*10^7$ seconds) the possible increase in thermal energy would be about $0.03$ joules. Trying to observe such an effect, if it even exists, would likely be futile. The rates for other macroscopic systems at manageable temperatures would not differ by more than a couple orders of magnitude. At ultra-cold temperatures the fractional increase would be far less.

What then are the prospects for experimental tests of equation \ref{2}? The technique of altering the basis in which interactions occur that is used in quantum eraser experiments is also a key element in the design of quantum computers. At the current stage of development environmental decoherence effects completely mask any inherent nonlinearities as small as those described earlier. Although it is not likely in the immediate future, one could hope that eventually the state of the art will  advance to a stage at which the predicted deviations might be observable.

The factor limiting the observability of collapse effects is the small ratio of interaction energy to $mc^2$. A relativistic version of equation \ref{2} would accommodate a much higher ratio, and also a more complete energy accounting, possibly offering additional experimental opportunities. To construct such an extension it is necessary to somehow reconcile the relativistic structure of spacetime with the nonlocal effects of quantum theory. The next section addresses this issue.

 \section{Toward a Relativistic Extension}

The fundamental principle shared by relativity and quantum theory is no-signaling. Given the nondeterministic nature of the connection between quantum theory and observation this principle must be understood as referring to the transmission of \textit{physical} instantiations of information. In \cite{Landauer} Landauer described this sort of information\footnote{This concept is distinct from more abstract, mathematical notions such as Fisher or Shannon information} as follows:
\begin{quote}
``Information is not an abstract entity but exists only through a physical representation, thus tying it to all the restrictions and possibilities of our real physical universe."
\end{quote}
Correlating interactions play the central role in the physical instantiation, transmission, and reproduction of information, and they are also responsible for generating the entanglement relations through which nonlocal effects are propagated. This is what motivated the hypothesis that is implemented in equation \ref {2}.

Given the nonlocal effects associated with entanglement the way that quantum field theory maintains the no-signaling principle is by assuming that spacelike-separated operators commute. This assumption of local commutativity implies the Born probability rule and, as pointed out by Weinberg\cite{Weinberg}(p.198), it is also guarantees the Lorentz invariance of the scattering matrix. In effect, the assumption of local commutativity is a third postulate of relativity. It supplements Einstein's original postulates of the equivalence of all inertial frames and the invariance of the speed of light. However, unlike Einstein's original postulates, it is unclear how it is connected to fundamental physical processes. The purpose of collapse equations is to establish such a connection. It would, of course, be preferable to do this while retaining the other two postulates.

The equivalence of inertial frames can be insured by replacing the preferred reference frame implied by equation \ref {2} with a randomly evolving spacelike hypersurface. The preferred frame can be viewed as just a special case of such a surface. The invariance of the speed of light can also be maintained. Incorporating these postulates into a collapse theory does introduce technical complications just as in the transition from nonrelativistic quantum mechanics to quantum field theory, but those issues will not be addressed here. Our current interest is in seeing how local commutativity emerges from fundamental interactions subject to a stochastic process, and in interpreting the additional structure that is attributed to relativistic spacetime.

Because the evolution of the spacelike surface is assumed to be random, 
and because the collapse effects that propagate along it are nondeterministic it remains hidden. So the sequence of spacelike-separated measurements is, in principle, unobservable, just as in conventional theory, and measurement outcomes are consistent with the Born probability rule. With this approach the postulate of local commutativity can be connected to fundamental processes. 

The light cone structure determined by the metric of relativistic spacetime would remain intact. To understand the randomly evolving surface that has been added to the metric it is helpful to recall Einstein's original Machian motivations in constructing both the special and general theories of relativity. In a 1921 address to the Prussian Academy of Sciences he stated:\cite{Geom_Exp}  \newline 
\begin{quote}
	``All practical geometry is based upon a principle which is accessible to
	experience..."
\end{quote}
He went on to explicitly acknowledge that the question of whether the relativistic structure of spacetime could be extended to the submolecular scale was one that must be answered by experiment. These statements reiterate his earlier views that properties attributed to space and time must be rooted in observations of physical objects 
\newline
\newline
\newline
\newline
\newline
and processes. Given the fact that the nondeterministic character of quantum theory implies that we cannot observe everything that happens at the most fundamental level it is reasonable to acknowledge that certain ontological features of spacetime could remain hidden from our view. The geometric structure that we attribute to spacetime reflects not only what we know, but also the physical limits of what is knowable. 

With this understanding we can see that an interaction based explanation of wave function collapse can provide a framework in which to construct a relativistic theory.

 \section{Discussion}

The assumption that wave function collapse is induced by correlating interactions leads to a stochastic collapse equation with several attractive properties. It  reduces to the Schr\"{o}dinger equation for stationary states and freely evolving systems, and deviates from it by a very small amount in situations that involve a few interacting elementary particles. It insures collapse to a definite outcome for systems of mesoscopic size with the correct probability and on a time scale that is consistent with our macroscopic experience. Because the strength and duration of the collapse effects are determined by the ratio of interaction energy to total relativistic energy it does not require the introduction of any new physical constants. 

Furthermore, it is consistent with exact conservation of momentum and orbital angular momentum in individual experiments, and also with energy conservation to the same extent that conventional nonrelativistic quantum theory correctly predicts such conservation. This consistency is based on the  recognition that all physical systems are quantum systems, and that they are almost always entangled to some extent. Since conserved quantities are shared across entangled branches the apparent nonconservation of a quantity in an elementary target system is compensated for by corresponding changes in (usually larger) systems with which it has interacted in the past.

Finally, compatibility with relativity can be achieved by replacing the fixed rest frame with an evolving spacelike surface. Nonlocal quantum effects propagate along the surface, and their nondeterministic nature prevents superluminal information transmission. This surface could be taken to evolve in a purely random fashion. This possibility respects the equivalence of all inertial frames, and maintains the light cone structure of relativistic spacetime, thus preserving the central principles of relativity. With this approach one might reasonably hope to develop an account that explains measurement outcomes in terms of fundamental processes, while preserving the essential features of contemporary theory.

\end{document}